\documentclass[showpacs,preprint]{revtex4-1}
\usepackage{amsmath}    
\usepackage{amssymb}
\usepackage{wasysym}
\usepackage{mathtools}
\usepackage{graphicx}   
\usepackage{verbatim}   
\usepackage{color}      
\usepackage{subfigure}  
\usepackage{hyperref}   
\usepackage{float}

\begin{document}

\title{Lovelock black hole thermodynamics in a string cloud model}
\author{Tae-Hun Lee$^{a,\;}$}
\email{taehunee@gmail.com}
\author{Sushant G. Ghosh$^{a,\;b\;}$}
\email{sgghosh@gmail.com}
\author{Sunil D. Maharaj$^{a\;}$}
\email{maharaj@ukzn.ac.za}
\author{Dharmanand Baboolal$^{a\;}$}
\email{Baboolald@ukzn.ac.za}

\affiliation{$^{a}$Astrophysics and Cosmology Research Unit, School of Mathematics, Statistics and Computer Science, University of KwaZulu-Natal, Private Bag X54001, Durban 4000, South Africa}
\affiliation{$^{b}$Centre for Theoretical Physics, Jamia Millia Islamia, New Delhi 110025, India}

\date{\today}

\begin{abstract}
The Lovelock theory is an extension of general relativity to higher dimensions. We study the Lovelock black hole for a string cloud model in arbitrary dimensional spacetime, and in turn also analyze its thermodynamical properties. Indeed, we compute the mass, temperature and entropy of the black hole and also perform a thermodynamical stability analysis. The phase structure suggests that the Hawking-Page phase transition is achievable.  It turns out that the presence of the  Lovelock terms and/or background string cloud completely changes the black hole thermodynamics. Interestingly, the entropy of a black hole is unaffected due to a background string cloud, but has a  correction term due to Lovelock gravity. 

\end{abstract}

\maketitle
\section{Introduction}
Amongst the higher curvature theories of gravity, Lovelock theory \cite{Lovelock:1971yv}, introduced by David Lovelock \cite{Lovelock:1971yv}, the action of which contains higher order curvature terms, is the most general second order gravity theory in higher dimensional spacetimes. 
Lovelock gravity \cite{Lovelock:1971yv} is a suitable and natural modification to higher dimensional scenarios in the sense that the additions to the Einstein action become divergence terms below a critical dimension, and hence they do not contribute to the equations of motion in the lower dimensions.  The Lovelock theories are exceptional, among the larger class of general higher curvature theories, in that the resulting field equations are just second derivatives of the metric function, and hence free from many pathologies that arise in higher curvature gravity. They are also shown to be free from ghosts when expanding about flat spacetime \cite{Wiltshire:1985us}. The Lovelock action goes back to the Einstein-Hilbert in four dimensions and contains corrections inspired by string theory to the Einstein-Hilbert action \cite{gsw}. The Lovelock gravities help us to explore in depth several conceptual issues of gravity in a much broader setup. Most interestingly, we can include features of black holes such as their existence and uniqueness theorems, their thermodynamical properties and their horizon structure, etc. Since their inception, significant attention has been given to black hole solutions, including their
formation, stability, and thermodynamics. 

The most extensive research done on the Einstein-Gauss-Bonnet gravity are with second order curvature corrections \cite{Boulware:1985wk, Wheeler:1985nh, Cai:2003kt, Dehghani:2005zm, Mora:2004kb, sgg}. It turns out that the Lagrangian and field equations are very complicated in the third order Lovelock gravity, but still there are a large number of works extending various black hole solutions to the third order Lovelock gravity \cite{Dehghani:2005zm, Mora:2004kb, Mazharimousavi:2008ti, Mazharimousavi:2009mb, Lee:2014dha, Ghosh:2014pga, Ghosh:2014dqa}. Recently, some works have been extended to general Lovelock gravity to find  the exact solutions and their thermodynamical properties  \cite{Mazharimousavi:2008ti}.  Although most of the known black hole solutions in Lovelock gravity are in vacuum, we may raise the question of having black hole solutions with  source. Hence, many authors have investigated the exact black hole solutions with source such as a string cloud background, for instance, in the Einstein-Gauss-Bonnet gravity \cite{Herscovich:2010vr,Mazharimousavi:2009mb}, and also in the third order Lovelock gravity \cite{Mazharimousavi:2009tp, Ghosh:2014pga, Ghosh:2014dqa}.

These days studying quantum effects of black holes is the most important route to a theory of quantum gravity. Hawking radiation \cite{Hawking:1974rv} is the most significant phenomenon and almost the only theoretical prediction for quantum effects in gravity so far. Furthermore, the Einstein theory itself has to be possibly modified to explain major astrophysical observations disagreeing with its predictions. In the development of physical theories, e.g. constructing quantum  gravities, unifying gravity with the rest of fundamental interactions or regularization schemes in the renormalization, etc, extending the spacetime dimension has often provided solutions for many problems. In such circumstances it is quite reasonable to consider modifications to the Einstein gravity in higher dimensions. On the other hand, string theory as one of strong candidates for quantum gravity and unified theories, contains such aspects. Whether or not string theory is an ultimate theory of the Universe, it will contribute to development of a final theory with its rich insights. In this sense to see effects of strings in modified gravity is a useful endeavor. Also, the universe can be represented as a collection of extended  objects and one-dimensional
strings, which is the most popular candidate for such fundamental objects.    It may be noted that  the study of Einstein's
equations coupled with a cloud of strings may be very important
because relativistic strings at a classical level can be used to
construct applicable models \cite{Letelier:1979ej}. The study of black holes
in string cloud models was pioneered by Letelier
\cite{Letelier:1979ej}  modifying the Schwarzschild solution for
a cloud of strings as a source \cite{Letelier:1979ej}, which was
recently extended to Gauss-Bonnet gravity
\cite{Herscovich:2010vr} and also to Lovelock gravity
\cite{Ghosh:2014dqa}. A string cloud background makes a profound influence
on horizon structure and thermodynamical quantities but the entropy is not
changed.

In this paper, we are concerned with the black
hole solutions of general Lovelock theory in the string cloud model, and we will try to understand how
 higher-curvature corrections due to Lovelock gravity and the background string 
change the qualitative thermodynamics of Lovelock black holes as we know from our experience with
black holes.  The examination of thermodynamical properties of black holes draws a great deal of attention to quantum gravity, and extensive studies in this direction are being done. Studying thermodynamics of Lovelock black holes was started by Myers and Simon \cite{Myers:1988ze}. Research involving the Lovelock black holes in the background of clouds of strings has been done \cite{Lee:2014dha, Ghosh:2014pga,Ghosh:2014dqa}. Here, we wish to also examine general thermodynamical properties of black holes in arbitrary dimensions for string cloud models.  

The paper is organized as follows: In Sec.~II  we begin by examining the Lovelock
action, which is a modification of the Einstein-Hilbert action, and
also derive energy momentum tensors for a cloud of strings. The
thermodynamics of a spherically symmetric Lovelock black hole solution in
this theory is examined in Sec.~III.  The thermodynamical stability is also discussed and we demonstrate that Hawking-Page phase transition is possible. 
Sec.~IV summarizes the results. In the appendix we discuss how locations of horizons and curvature singular points are affected by changing parameters, especially the mass parameter. We used units that fix $G=c=1$ and the metric
signature, $(-,+,+,\cdots,+)$.

\section{Lovelock Black Holes}
Lovelock gravity is the most general higher dimensional second order theory of gravity which is free of ghosts. Lovelock gravity action in $N(\equiv n+2)$ dimensions reads \cite{Lovelock:1971yv}
\begin{equation}
\mathcal{I}=\int dx^{n+2}\sqrt{-g}\sum\limits_{s=0}^{m}\alpha _{s}\mathcal{L}_{s}+\mathcal{I}_{matter},\label{action}
\end{equation} 
where $\alpha_s$ is an arbitrary constant, $\mathcal{L}_s$ is the Euler density and the generalized Kronecker delta $\delta _{\alpha _{1}\beta_{1}...
\alpha _{s}\beta _{s}}^{\mu _{1}\nu _{1}...\mu _{s}\nu_{s}}$ is totally antisymmetric in both sets of indices. 
\begin{equation}
\mathcal{L}_s=\frac{1}{2^{s}}\delta _{\alpha _{1}\beta_{1}...
\alpha _{s}\beta _{s}}^{\mu _{1}\nu _{1}...\mu _{s}\nu_{s}}
\prod\limits_{r=1}^{s}R_{\quad \mu _{r}\nu _{r}}^{\alpha _{r}\beta _{r}},\label{EulerDensity}
\end{equation}
and $2m=n$ for even dimensions and $2m-1=n$ for odd dimensions. Here the Lovelock tensor is nonlinear in the Riemann tensor and differs from the Einstein tensor in $N=n+2\geq 4$. We choose $\mathcal{L}_0=1$ and $\alpha_0=\Lambda$. General relativity is a special case of Eq.~(\ref{EulerDensity}) where $\mathcal{L}_1=R$ and only $\alpha_1$ is nonvanishing.
Here we want to study $N-$dimensional static spherical black holes for a string cloud model in general Lovelock theory. The Einstein-Lovelock equations read (assuming $\alpha_0=0$)
\begin{equation}
G^{\nu(1)}_\mu+\sum\limits^{m}_{s=2}\alpha_s G^{\nu(s)}_\mu=T^\nu_\mu. \label{Einstein eq}
\end{equation}
$G^{\nu(1)}_\mu$ is the Einstein tensor, where the second terms are due to Lovelock gravity, and we do not give their expression.
Here we wish to consider the energy momentum tensor of a cloud of strings, which can be obtained from the Nambu-Goto action of a string \cite{Letelier:1979ej},
\begin{equation}
\mathcal{I}_S = m\int_{\Sigma}  (-\gamma)^{1/2} \;  d\lambda^{0} d\lambda^{1},
\end{equation}
where $m>0$ is a positive constant for each string and $\gamma=\det{\gamma_{a b}}$ where
\begin{equation}
 \gamma_{a b} = g_{\mu \nu} \frac{\partial x^{\mu}}{\partial \lambda^{a}} \frac{\partial x^{\nu}}{\partial \lambda^{b}},
\end{equation}
where $(\lambda^{0}, \lambda^{1})$ with $\lambda^{0}$ and $\lambda^{1}$ being timelike and spacelike parameters. Using $T^{\mu \nu} = -2 \partial \mathcal{L}/\partial g^{\mu \nu}$, the energy momentum tensor for a cloud of strings reads
\begin{equation}
T^{\mu \nu} = \rho {\Sigma^{\mu \sigma} \Sigma_{\sigma}^{\phantom{\sigma} \nu}}/{(-\gamma)^{1/2}  },\label{EM}
\end{equation}
where
\begin{equation}
\Sigma^{\mu \nu} = \epsilon^{a b} \frac{\partial x^{\mu}}{\partial \lambda^{a}} \frac{\partial x^{\nu}}{\partial \lambda^{b}},
\end{equation}
$\epsilon^{a b}$ is the Levi-Civita tensor and $\rho$ is a proper density of a cloud of strings.
We assume a metric in the following form \cite{Myers:1988ze, Wheeler:1985nh, Wheeler:1985qd, Lee:2014dha}
\begin{equation}\label{metric}
ds^2=-f(r) dt^2+ \frac{1}{f(r)} dr^2 + r^2 \tilde{\gamma}_{ij}\; dx^i\; dx^j,
\end{equation}
where $\tilde{\gamma}_{ij}$ is a metric in the $n$-dimensional constant curvature space ($\kappa = 1, 0, -1$). It has been shown by Wheeler \cite{Wheeler:1985qd} that equations of motion for general vacuum solutions of the Lovelock action in static spherically symmetric spacetimes are obtained by solving a polynomial equation. We assume
\begin{equation}
f(r)=\kappa-r^2 F(r),\label{ftoF}
\end{equation}
and then the $tt$ component of Eq.~(\ref{Einstein eq}) reads
\begin{equation}
\sum\limits^m_{s=0}\tilde{\alpha}_sF^s=\frac{\omega_0} {r^{n+1}}-\frac{V(r)}{r^{n+1}},\label{polysol}
\end{equation}
with $\tilde{\alpha}_s$ by
\begin{equation}
\begin{array}{cc}
&(n+1)n\tilde{\alpha}_0=\alpha_0,\\
&\tilde{\alpha}_1=1,\\
&\tilde{\alpha}_s=\prod\limits^{2s}_{i=3}(n+1-i)\alpha_s~~\text{for}~s>1.
\end{array}
\end{equation}
Here $\omega_0$ is an integration constant related to the Arnowitt-Deser-Misner(ADM) mass defined by 
\begin{equation}
M=\xi_n\omega_0=\frac{n}{8\pi}\frac{\pi^{(n+1)/2}}{\Gamma[(n+1)/2]}\omega_0,
\end{equation}
and $V(r)$ is related to the energy momentum tensor, explicitly written as
\begin{equation}
V(r)=\frac{2}{n}\int  r^nT^t_t dr.
\end{equation}
The general form of the energy momentum tensor $T^\mu_\nu$ for static spherically symmetric spacetimes \cite{{Mazharimousavi:2008ti}} is given by
\begin{equation}
T^\mu_{\nu}=\frac{a}{r^{n(1-k)}}\text{diag}(1,1,k,\cdots, k),\label{energy_momentum}
\end{equation}
where $k$ is a constant.
The dominant energy condition allows only $a\leq0$ and $-1\leq k\leq0$ and the causality condition further constrains $k$ to $-1\leq k\leq-1/n$ \cite{Mazharimousavi:2008ti}. In a sense the string cloud is unique in that its energy momentum tensor corresponds to $k=0$, which contributes to the mass with the highest power. General expressions of thermodynamical quantities can be conveniently obtained using the polynomial equation. Reference \cite{Cai:2003kt} gives thermodynamical quantities for static spherically symmetric vacuum solutions. Henceforth, we specialize to the case of a string cloud model, i.e., $k=0$.
To express $f(r)$ in terms of $r$ we need to solve the above polynomial equation. For a special choice of $\tilde{\alpha}_s$ $f(r)$ can be expressed explicitly. For instance, given the special relation $\tilde{\alpha}_s=(\pm 1)^{s+1}\left(\frac{[(n+1)/2]}{s}\right)\ell^{2s-n-2}$ and $\tilde{\alpha}_0/\tilde{\alpha}_1=\pm(\ell^{-2}/[n+1/2])$, $f(r)$ for even $n$ and odd $[(n+1)/2]$, where $[\cdot]$ refers to the integer part, can be written as \cite{Mazharimousavi:2008ti}
\begin{equation}
f(r)=\kappa\pm\frac{r^2}{\ell^2}\mp\frac{r^2}{\ell^2}\left(\pm\frac{n}{2}\frac{1}{\ell^{n-2}}\times\left\{
\begin{array}{lll}
&\frac{M}{r}-\frac{2a}{n(nk+1)}r^{nk},&~nk+1\neq 0\\
&\frac{M}{r}-\frac{2a}{n}\frac{\ln r}{r},&~nk+1=0
\end{array}
\right.\right)^{2/n}.
\end{equation}

\begin{figure}[H]
  \begin{center}
  \begin{tabular}{lll}
     \includegraphics[scale=0.7]{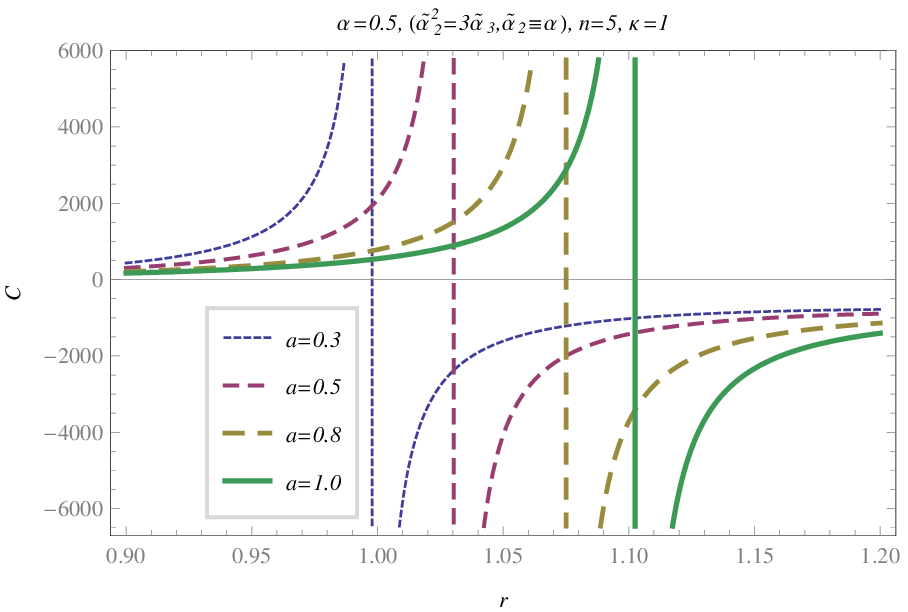} &&
    \includegraphics[scale=0.7]{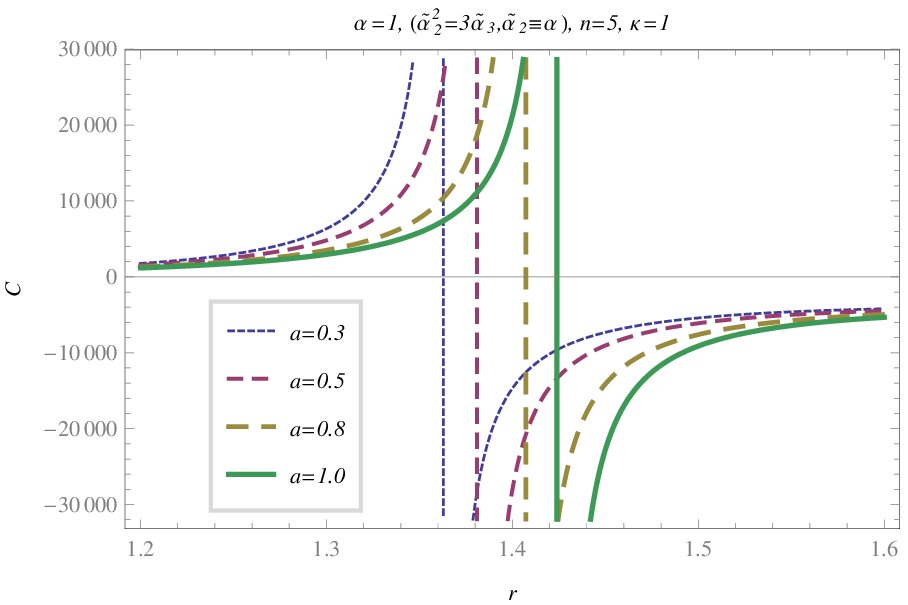}
\end{tabular} 
    \end{center}
    \caption{Heat capacity in $n=5$ for $\tilde{\alpha}_0=0, \tilde{\alpha}^2_2=3\tilde{\alpha}_3, \tilde{\alpha}_2=\alpha=0.5, 1.0$ and $\kappa=1$}\label{}
\end{figure}
\begin{figure}[H]
  \begin{center}
  \begin{tabular}{ccc}
     \includegraphics[scale=0.8]{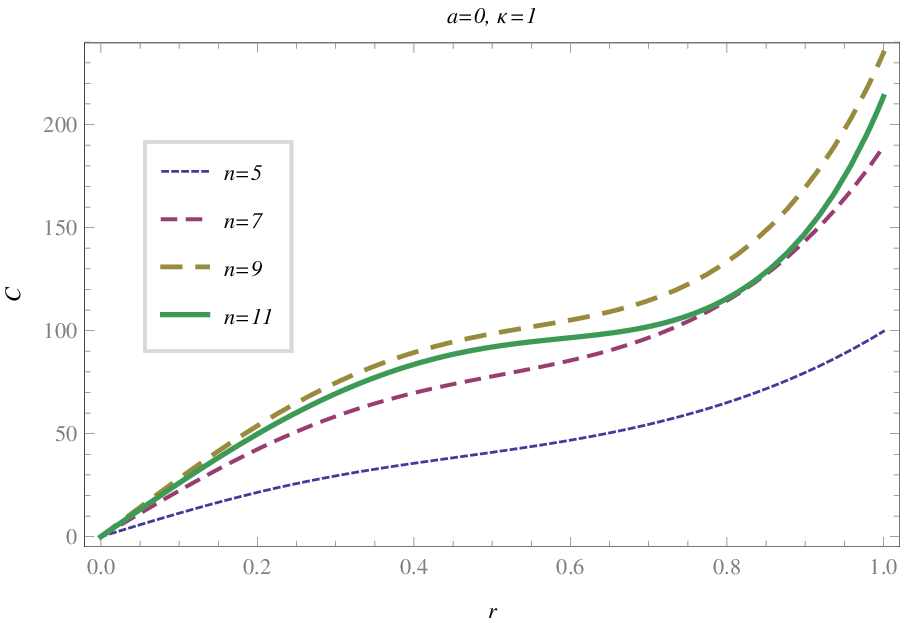} &&
    \includegraphics[scale=0.8]{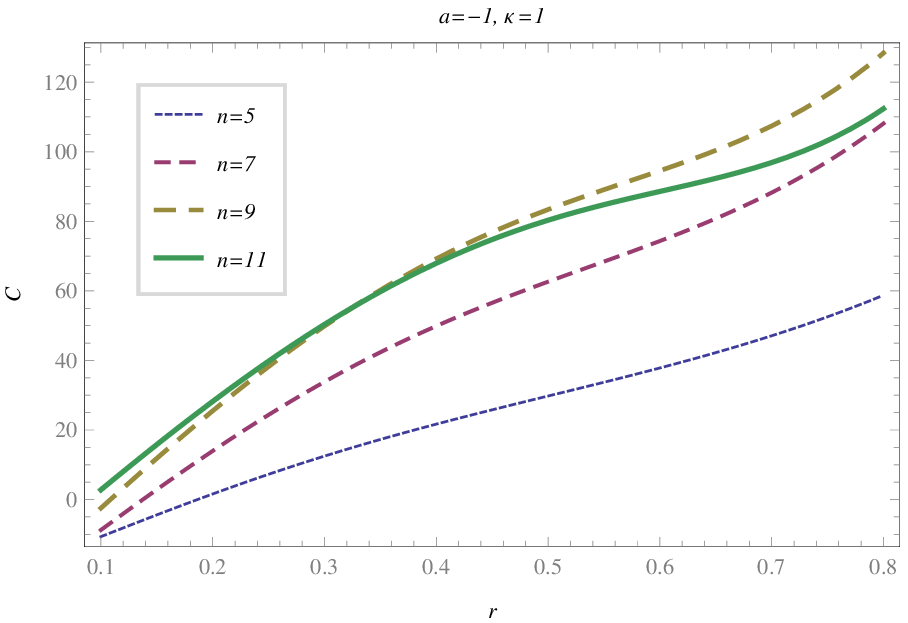}
\end{tabular} 
    \end{center}
    \caption{Heat capacity for $a=0,-1$, $\tilde{\alpha}_i=1$ and $\kappa=1$ in odd deminsions. In this case there exists no critical point}\label{}
\end{figure}
\begin{figure}[H]
  \begin{center}
  \begin{tabular}{ccc}
     \includegraphics[scale=0.8]{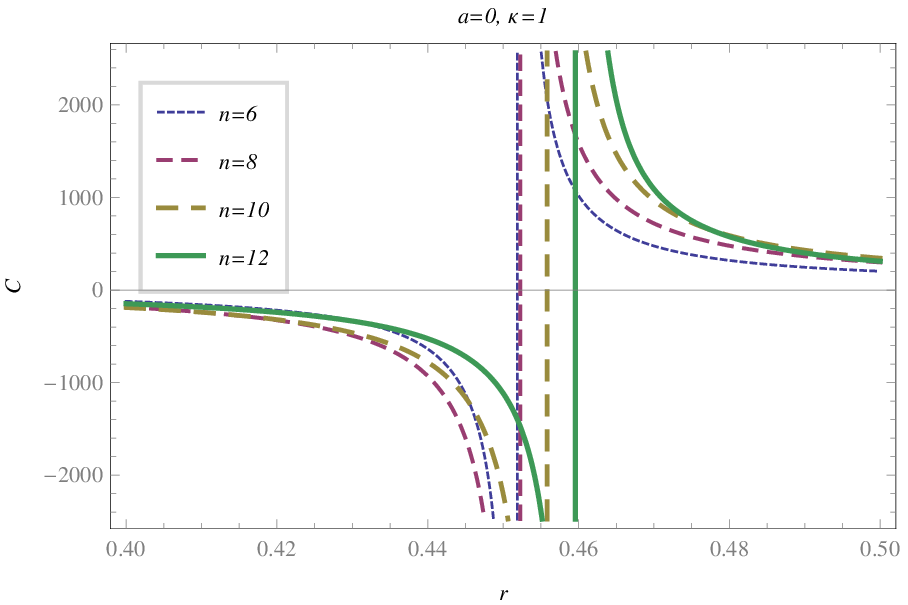} &&
    \includegraphics[scale=0.8]{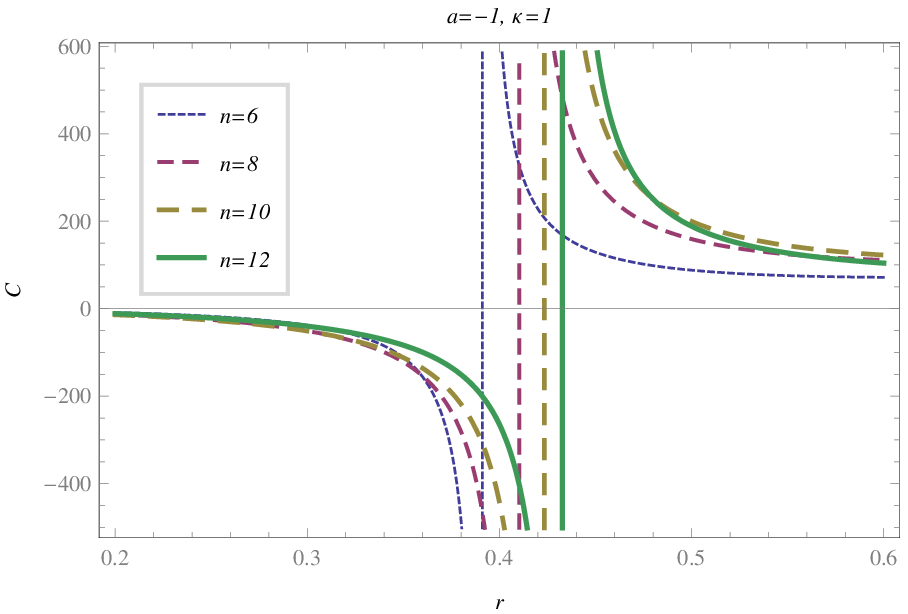}
\end{tabular} 
    \end{center}
    \caption{Heat capacity for $a=0,-1$, $\tilde{\alpha}_i=1$ and $\kappa=1$ in even deminsions. In this case there exists a critical point}\label{}
\end{figure}
\begin{figure}[H]
  \begin{center}
  \begin{tabular}{ccc}
     \includegraphics[scale=0.8]{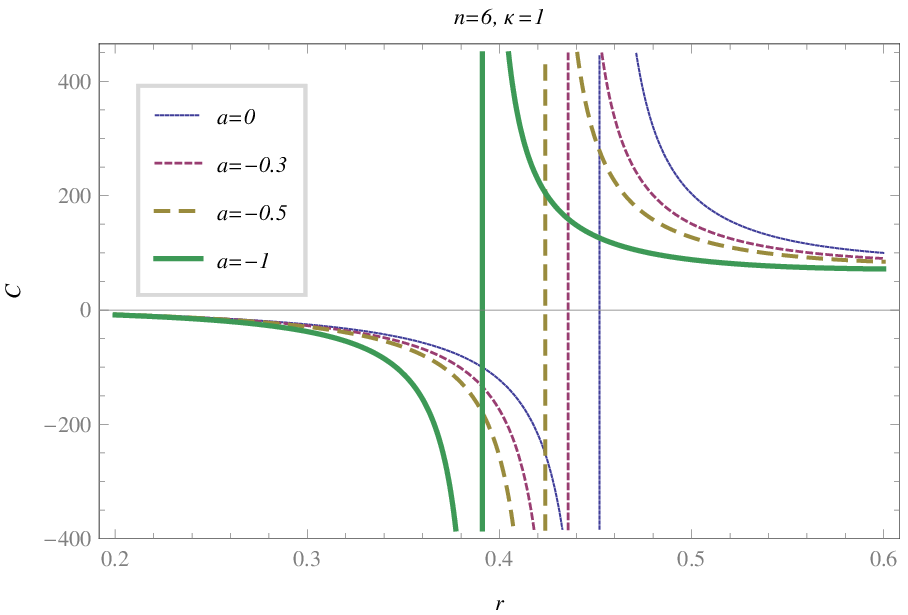} &&
    \includegraphics[scale=0.8]{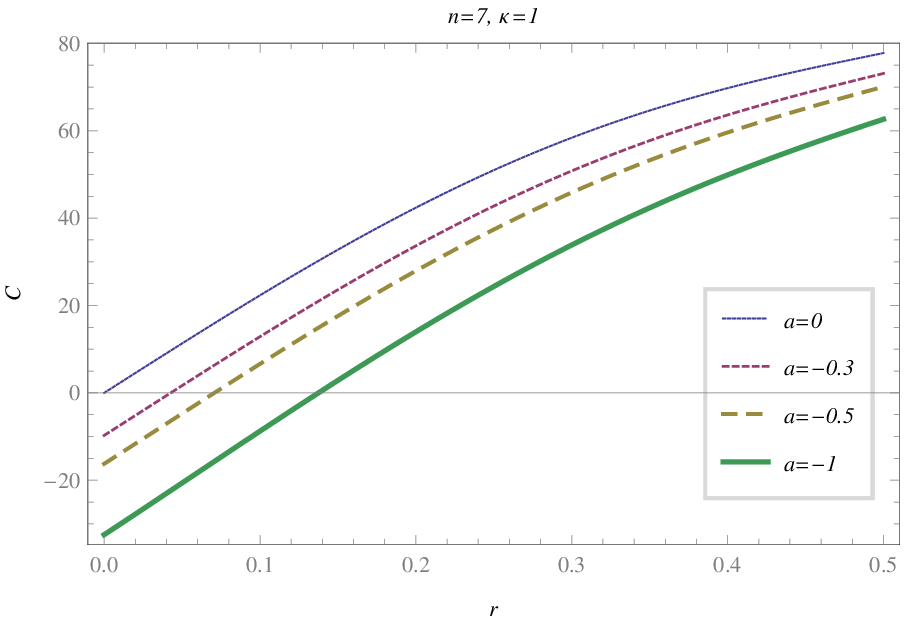}
\end{tabular} 
    \end{center}
    \caption{Plots showing heat capacity for $\tilde{\alpha}_i=1$ and $\kappa=1$ in $N=8$ and $N=9$ deminsions as $a$ changes.}\label{}
\end{figure}

\section{Thermodynamics and Phase Transition}
In this section, we present the exact thermodynamical quantities associated with the Lovelock black solution in Eq.~(\ref{polysol}). Based on the general results in the last section we find general thermodynamical quantities in terms of a horizon radius. Let us introduce a new function $\omega$ such that
\begin{equation}
\omega(r,f)=\sum\limits^m_{s=0}\tilde{\alpha}_s(\kappa-f)^sr^{-2s+n+1}+V(r),
\end{equation}
with $\omega_0=\omega(r_h,0)$. The Arnowitt-Deser-Misner(ADM) mass is defined by 
\begin{equation}
M=\frac{nV_n}{16\pi}\omega_0,
\end{equation}
where $V_n=2\pi^{(n+1)/2}\Gamma[(n+1)/2]$ and using Eq.~(\ref{polysol}) we get
\begin{equation}
M=\xi_n\sum\limits^m_{s=0}\tilde{\alpha}_s\kappa^sr^{-2s+n+1}_h+\xi_n V(r_h).\label{M}
\end{equation}
For $V(r)=0$ mass reduces to the result in \cite{Cai:2003kt}, 
\begin{equation}
M=\xi_n\sum\limits^m_{s=0}\tilde{\alpha}_s\kappa^sr^{-2s+n+1}_h.\label{Ma0}
\end{equation}
For $\kappa=0$
\begin{equation}
M=\xi_n\tilde{\alpha}_0 r^{n+1}_h+\xi_n V(r_h).
\end{equation}
For $f$ explicitly given as a function of $r$ and a constant $\omega$,  the Hawking temperature $T_h$ for a black hole is defined by $T_h=\kappa_s/2\pi$, where $\kappa_s$ is a surface gravity and hence is expressed as   
\begin{equation}
T_h=\frac{1}{4\pi}\left.\frac{\partial f}{\partial r}\right|_{r=r_h}.
\end{equation}
The quantity $\left.\partial f/\partial r\right|_{r=r_h}$ in the definition can be interpreted as a velocity $df/dr$ at a position $f=0$ on the line $\omega(r,f)=\text{constant}$. That is, from the vanishing total derivative $\left.d\omega/dr\right|_{f=0}=0$ for the function $\omega(r,f)$, we can write the Hawking temperature as
\begin{equation}
T_h=\frac{1}{4\pi}\left.\frac{df}{dr}\right|_{f=0}=-\frac{1}{4\pi}\frac{\partial\omega}{\partial r}\left.\frac{\partial\omega}{\partial f}\right|_{f=0}.
\end{equation}
Using derivatives of $\omega$ with respect to $f$ and $r$,
\begin{eqnarray}
\frac{\partial\omega}{\partial f}&=&-\sum\limits^m_{s=1}\tilde{\alpha}_ss(\kappa-f)^{s-1}r^{-2s+n+1},\nonumber\\
\frac{\partial\omega}{\partial r}&=&\sum\limits^m_{s=0}\tilde{\alpha}_s(\kappa-f)^{s}(-2s+n+1)r^{-2s+n}+\frac{\partial V}{\partial r},\label{dwdfdr}
\end{eqnarray}
we obtain the Hawking temperature
\begin{equation}
T_h=\frac{1}{4\pi}\frac{\sum\limits^m_{s=0}\tilde{\alpha}_s\kappa^{s}(-2s+n+1)r^{-2s+n}_h+\frac{\partial V}{\partial r_h}}{\sum\limits^m_{s=1}\tilde{\alpha}_s s\kappa^{s-1}r^{-2s+n+1}_h}.\label{T}
\end{equation}
$T_h$ diverges at a curvature singular point $r_h=r_*$. Without a source term $V(r)=0$,  $T_h$ becomes the expression \cite{Cai:2003kt}  
\begin{equation}
T_h=\frac{1}{4\pi}\frac{\sum\limits^m_{s=0}\tilde{\alpha}_s\kappa^{s}(-2s+n+1)r^{-2s+n}_h}{\sum\limits^m_{s=1}\tilde{\alpha}_s s\kappa^{s-1}r^{-2s+n+1}_h}.\label{Ta0}
\end{equation}
 In the following discussion let us assume $r_h$ is the greatest horizon radius and $r_*$ is the greatest singular point, and for simplicity take $\kappa=1$. If the highest order coefficient is positive, e.g. $\tilde{\alpha}_0>0$, then the slope at the greatest zero is nonnegative, i.e. $\left.\partial\Omega/\partial r\right|_{r_h}=\left.\partial\omega/\partial r\right|_{f=0, r_{h}}\geq0$. On the other hand, we can see that due to the asymptotic flatness of $f(r)$ as $r$ goes to infinity, the quantity $-\partial\omega/\partial f$ indicating a curvature singular point asymptotically goes to infinity for $\tilde{\alpha}_1>0$. Thus, in the region $r_{h}>r_{*}$, $-\left.\partial\omega/\partial f\right|_{f=0,r_h}\geq0$. Under these conditions, $T_h\geq0$. Roots can be found in the expression $\Omega(r_h)=0$ having $\left.\partial\omega/\partial r\right|_{r_h}<0$. However, those regions do not correspond to the greatest horizon. It is also worthwhile to notice that in the limit $r_{h}\to r_{*}$, $T_h\to+\infty$ and as $r_h\to\infty$, $T_h\to\infty$ for $\tilde{\alpha}_0\neq0$ and $T_h\to0$ for $\tilde{\alpha}_0=0$. This means that between $r_{*}$ and $\infty$, if there exists a region where $\partial T_h/\partial r_h$ is positive, then there must be points where $\partial T_h/\partial r_h=0$.
Entropy can also be expressed in a simple way and it can be easily seen that the condition $r_{h}>r_{*}$ guarantees positivity of entropy. Using the first law of thermodynamics, $dM=TdS$,
\begin{equation}
dS=\frac{dM}{T}dr=\xi_n\frac{dr}{T}\left.\frac{\partial\omega}{\partial r}\right|_{f=0}=-4\pi\xi_n dr\frac{\partial\omega}{\partial r}\frac{\partial\omega}{\partial f}\Big/\left.\frac{\partial\omega}{\partial r}\right|_{f=0}=-4\pi\xi_n dr\left.\frac{\partial\omega}{\partial f}\right|_{f=0}.
\end{equation}
From Eq.~(\ref{dwdfdr}) we see that it is a general result that entropy should not depend on the energy momentum tensor at all. This result coincides with the corresponding case in vacuum \cite{Cai:2003kt}. By integration we get the entropy
\begin{equation}
S=\int^{r_h}_0dS=4\pi\xi_n\int^{r_h}_0\sum\limits^m_{s=1}\tilde{\alpha}_ss\kappa^{s-1}r^{-2s+n+1}dr
=4\pi\xi_n\sum\limits^m_{s=1}\frac{\tilde{\alpha}_s\kappa^{s-1}s}{-2s+n+2}r^{-2s+n+2}_h,\label{S}
\end{equation}
where an arbitrary additive constant has been omitted. With the above discussion $dS>0$ for any $r>0$ implies that a curvature singular point $r_*$ only exists within a horizon radius $r_h$.
Heat capacity is used in the following expression,
\begin{equation}
C=\frac{\partial M}{\partial T}=\frac{\partial M}{\partial r_h}\Big/\frac{\partial T_h}{\partial r_h}=T_h\frac{\partial S}{\partial T_h}=T_h\frac{\partial S}{\partial r_h}\Big/\frac{\partial T_h}{\partial r_h}.\label{Ct}
\end{equation}
For $r_{h}>r_{*}$, $dS>0$, i.e. $\partial S/\partial r_h>0$ and  $T_h>0$. For $T_h>0$ the sign of $C$ depends only on $\partial T_h/\partial r_h$. There exists an even number of transition points $r_{c}$ after $r_*$ such that $\left.\partial T_h/\partial r_h\right|_{r=r_{c}}=0$ where the heat capacity is divergent. If $r_{cM}$ is the greatest $r_c$, then for $r_h>r_{cM}$, $\partial T_h/\partial r_h<0$ and heat capacity is negative (but positive for $r_h<r_{cM}$). Therefore, it is a general result using $\tilde{\alpha}_0>0$ and $\tilde{\alpha}_1>0$, that black holes are not thermodynamically stable for $r_h>r_{cM}$. The heat capacity is explicitly given by
\begin{equation}
C=4\pi\xi_n\frac{AB^2}{D},\label{C}
\end{equation}
where 
\begin{equation}
\begin{array}{ccl}
A&=&\left.\frac{\partial\omega}{\partial r}\right|_{f=0}=\sum\limits^m_{s=0}\tilde{\alpha}_s\kappa^{s}(-2s+n+1)r^{-2s+n}_h+\frac{\partial V}{\partial r_h},\\
B&=&-\left.\frac{\partial\omega}{\partial f}\right|_{f=0}=\sum\limits^m_{s=1}\tilde{\alpha}_s s\kappa^{s-1}r^{-2s+n+1}_h,\\
D&=&B\partial_{r_h}A-A\partial_{r_h}B,
\end{array}
\end{equation}
and
\begin{equation}
\begin{array}{ccl}
D&=&\sum\limits^m_{s,p}\tilde{\alpha}_s\tilde{\alpha}_p\kappa^{s+p-1}s(-2p+2s-1)(-2p+n+1)r^{-2(s+p-n)}_h\\
&&-\partial_{r_{h}}V\sum\limits^m_{s=1}\tilde{\alpha}_ss\kappa^{s-1}(-2s+n+1)r^{-2s+n}_h+\partial^2_{r_{h}}V\sum\limits^m_{s=1}\tilde{\alpha}_ss\kappa^{s-1}r^{-2s+n+1}_h.
\end{array}
\end{equation}
When there is no source term, heat capacity is reduced to the expression in \cite{Cai:2003kt}:
\begin{equation}
C=4\pi\xi_n\frac{\left[\sum\limits^m_{s=0}\tilde{\alpha}_s\kappa^{s}(-2s+n+1)r^{-2s+n}_h\right]\left[\sum\limits^m_{s=1}\tilde{\alpha}_s s\kappa^{s-1}r^{-2s+n+1}_h\right]^2}{\sum\limits^m_{s,p}\tilde{\alpha}_s\tilde{\alpha}_p\kappa^{s+p-1}s(-2p+2s-1)(-2p+n+1)r^{-2(s+p-n)}_h}.\label{Ca0}
\end{equation}
We are particularly interested in a cloud of strings as a source of matter here. Thermodynamics of black holes in a string cloud background has been studied in the Gauss-Bonnet gravity \cite{Herscovich:2010vr}. The action and the energy momentum tensors for static spherically symmetric spacetimes due to cloud of strings have been obtained in \cite{Letelier:1979ej}. The general energy momentum tensor for static spherically symmetric spacetimes is given in \cite{Mazharimousavi:2008ti}. In Eq.~(\ref{energy_momentum}),  $k=0$ corresponds to a cloud of strings \cite{Herscovich:2010vr}. Thus, $V(r)$ is identified with
\begin{equation}
V(r)=\frac{2a}{n}r.
\end{equation}
The case $k=0$ gives the upper bound satisfying the dominant energy condition (the causality condition lowers the upper bound down to $-1/n$.)\cite{Mazharimousavi:2008ti}. This means a cloud of strings contributes to the mass in the highest coupling with the highest possible order for the energy momentum tensor in the horizon radius. From Eq.~(\ref{M}) the mass of the black hole is given 
\begin{equation}
M=\xi_n\sum\limits^m_{s=0}\tilde{\alpha}_s\kappa^sr^{-2s+n+1}_h+\xi_n\frac{2a}{n}r_h.\label{Ms}
\end{equation}
We  notice that due to the $1/n$ factor in the potential for large dimensions the effect of the source for thermodynamic quantities gets small. 
For odd dimensions a string cloud linear contribution is unique while for even dimensions, since it can be combined with $\tilde{\alpha}_m\kappa^m r^{-2m+n+1}_h$ by redefining the highest coupling $\tilde{\alpha}^\prime_m=\tilde{\alpha}_m+a/(m\kappa^m)$, the change of the coupling $a$ is effectively that of the value of $\tilde{\alpha}_m$.
We see that the contribution of strings to mass linearly depends on $r_h$. Choosing $n=5$ and $\tilde{\alpha}^2_2=3\tilde{\alpha}_3$ gives 
\begin{equation}
M=\frac{\pi^2}{16}\left(2ar_h+\frac{5}{3}\kappa^3\alpha^2+5\kappa r_h^4+5\kappa^2\alpha r^2_h\right),
\end{equation}
where $\tilde{\alpha}_2\equiv\alpha$. This result can be verified in \cite{Lee:2014dha} for $\kappa=1$. The Hawking temperature is
\begin{equation}
T_h=\frac{1}{4\pi}\frac{\sum\limits^m_{s=0}\tilde{\alpha}_s\kappa^{s}(-2s+n+1)r^{-2s+n}_h+\frac{2a}{n}}{\sum\limits^m_{s=1}\tilde{\alpha}_s s\kappa^{s-1}r^{-2s+n+1}_h}.\label{Ts}
\end{equation} 
As we found before, entropy does not depend on the matter action
\begin{equation} 
S=4\pi\xi_n\sum\limits^m_{s=1}\tilde{\alpha}_s\kappa^{s-1}\frac{s}{-2s+n+2}r^{-2s+n+2}_h.\label{Ss}
\end{equation}
For $\kappa=0$ the entropy satisfies the area law.
\begin{equation} 
S=\frac{n}{4}\frac{2\pi^{(n+1)/2}}{\Gamma[(n+1)/2]}r^{n}_h=\frac{A}{4}.
\end{equation}
For $n=5$ and $\tilde{\alpha}^2_2=3\tilde{\alpha}_3$, 
\begin{equation}
T_h=\frac{1}{4\pi}\frac{4\kappa r_h^3+2\kappa^2\alpha r_h+2a/5}{(r^2_h+\kappa\alpha)^2},
\end{equation}
and
\begin{equation}
S=\frac{\pi^3}{12}\left(3r^5_h+10\kappa\alpha r^3_h+15\kappa^2\alpha^2 r_h\right),
\end{equation}
which were found for $\kappa=1$ in \cite{Lee:2014dha}. 
For $\kappa=0$
The heat capacity is written as
\begin{equation}
C=4\pi\xi_n\frac{AB^2}{D},\label{C}
\end{equation}
where
\begin{equation}
\begin{array}{ccl}
A&=&\left.\frac{\partial\omega}{\partial r}\right|_{f=0}=\sum\limits^m_{s=0}\tilde{\alpha}_s\kappa^{s}(-2s+n+1)r^{-2s+n}_h+\frac{2a}{n},\\
B&=&-\left.\frac{\partial\omega}{\partial f}\right|_{f=0}=\sum\limits^m_{s=1}\tilde{\alpha}_s s\kappa^{s-1}r^{-2s+n+1}_h,
\end{array}
\end{equation}
and
\begin{equation}
\begin{array}{ccl}
D&=&\sum\limits^m_{s,p}\tilde{\alpha}_s\tilde{\alpha}_p\kappa^{s+p-1}s(-2p+2s-1)(-2p+n+1)r^{-2(s+p-n)}_h\\
&&-\frac{2a}{n}\sum\limits^m_{s=1}\tilde{\alpha}_ss\kappa^{s-1}(-2s+n+1)r^{-2s+n}_h.
\end{array}
\end{equation}
For $n=5$ and $\tilde{\alpha}^2_2=3\tilde{\alpha}_3$,
\begin{equation}
\begin{array}{cc}
&A=2(2 \kappa r^3_h+\kappa^2\alpha r_h)+2a/5,~~B=(r^2_h+\kappa\alpha)^2\\
&D=-2\kappa(r^2_h+\kappa\alpha)(2 r^4_h-3\kappa\alpha r^2_h-\kappa^2\alpha^2)-\frac{8a}{5}r_h(r^2_h+\kappa\alpha).
\end{array}
\end{equation}
Hence ,
\begin{equation}
C=-\frac{5}{4}\pi^3\frac{[5(2\kappa r^3_h+\kappa^2\alpha r_h)+a](r^2_h+\kappa\alpha)^3}{5\kappa(-\kappa^2\alpha^2+2r^4_h-3\kappa\alpha r^2_h)+4a r_h}.
\end{equation}
This result was established  in \cite{Lee:2014dha} for $\kappa=1$ verifying the general expression in Eq.~(\ref{C}).
It can be easily noticed that for $a=0$ all thermodynamic quantities reduce to those in \cite{Cai:2003kt}. To see a non-trivial contribution of matter to heat capacity we consider $T_h=0$ and $\partial T_h/\partial r_h$ in the expression of heat capacity in Eq.~(\ref{Ct}). If $T_h=0$ exists somewhere in $r_*<r_h<\infty$ then a zero temperature horizon $r_{hT}$ can be found where
\begin{equation}
\sum\limits^m_{s=0}\tilde{\alpha}_s\kappa^{s}(-2s+n+1)r^{-2s+n}_{hT}+\frac{2a}{n}=0.
\end{equation}
If $\alpha_0>0$, then a string cloud will increase $r_{hT}$ because $2a/n<0$. For $\partial T_h/\partial r_h=0$, we write 
\begin{equation}
\frac{\partial T_h}{\partial r_h}=\frac{\partial T_{h0}}{\partial r_h}+\frac{(-2a/n)}{B^2}\frac{\partial B}{\partial r_h},
\end{equation}
where $T_{h0}$ is the Hawking temperature without a source, and $B=\sum\limits^m_{s=1}\tilde{\alpha}_s s\kappa^{s-1}r^{-2s+n+1}_h$ is a denominator in the Hawking temperature, which identifies a curvature singular point $r_*$.
Asymptotically $\partial T_{h0}/\partial r_h<0$ at infinity the value for $\tilde{\alpha}_0, \tilde{\alpha}_1>0$. $\partial B/\partial r_h>0$ for $r_h>r_*$ if $r_*$ is the greatest zero of $B$, so $(-2a/nB^2)\partial B/\partial r_h>0$. Hence $r_c$ becomes greater and the thermodynamically stable region, if it exists, is decreased by a string effect.

\section{Summary}
We have studied static and spherically symmetric black hole solutions in Lovelock gravity in arbitrary $N=n+2$ dimensions with an energy momentum tensor given by a cloud of strings. Thus, we have generalized the previous studies of black hole solutions in a string cloud background. We performed a detailed analysis of thermodynamical aspects of Lovelock black holes and the impact of the string cloud background. Although the geometry and horizon structures of the black hole solutions are quite complicated, we found the exact expression for thermodynamical quantities like the black hole mass, the Hawking temperature, entropy and heat capacity. We have explicitly shown the effect of a string cloud on black hole thermodynamics. Interestingly, it turns out that we can find a set parameters for which black holes are thermodynamically stable and the Hawking-Page phase transition is achievable. The entropy of a black hole is unaffected by a background string cloud. In classical general relativity, the black hole entropy obeys the area law, whereas in the Lovelock black holes considered here the area law is not changed.

\appendix

\section{Range of Parameters for a black hole}
If we require that a physically acceptable black hole should not reveal a curvature singular point outside the horizon, then there exists an allowed domain of a parameter set. A curvature singular point stays not only at a centre but also outside the black hole. For instance, the reference \cite{Myers:1988ze} discusses such a singularity in Gauss-Bonnet gravity. We can extend our analysis to general Lovelock gravity.   
As mentioned in the last section obtaining a metric function $f(r)$ is reduced to solving a polynomial equation for $F(r)$. Therefore, $F(r)$ and hence $f(r)$ contain terms having a fractional exponent. At points where such terms vanish, derivatives of $f(r)$ and hence curvature terms can diverge, \cite{Myers:1988ze}. For instance, $f=A+B^{1/p}$, where $p$ is an integer and $A$ and $B$ are analytic functions and $B$ is not an integer exponentiation of an analytic function. $\partial f/\partial r$ diverges at $B=0$ as long as $AB^{1-1/p}+B^\prime/p\neq0$. As we will see later it is an important criterion for physical properties of thermodynamical quantities and determining whether such singular points stay inside a horizon. Let us define a function identifying a singular point as a denominator of $d f/d r$:
\begin{equation}
\frac{d f}{d r}=-\frac{\partial\Omega/\partial r}{\partial\Omega/\partial f}=\frac{\sum\limits^m_{s=0}\tilde{\alpha}_s(\kappa-f)^s(-2s+n+1)r^{-2s+n}+V^\prime}{\sum\limits^m_{s=1}\tilde{\alpha}_ss(\kappa-f)^{s-1}r^{-2s+n+1}},
\end{equation}
obtained by using the relation Eq.~(\ref{ftoF}) and taking the derivative of Eq.~(\ref{polysol}). Note that $\Omega$ is defined by
\begin{equation}
\Omega\equiv\sum\limits^m_{s=0}\tilde{\alpha}_s(\kappa-f)^sr^{-2s+n+1}+V-\omega_0.\label{Omega}
\end{equation}
Then a curvature singular point $r_*$ and a horizon radius $r_h$ can be obtained from
\begin{equation}
\left.\frac{\partial \Omega}{\partial f}\right|_{r_*}=-\sum\limits^m_{s=1}\tilde{\alpha}_ss[\kappa-f(r_*)]^{s-1}r^{-2s+n+1}_*=0, \label{eq:r*}
\end{equation}
and
\begin{equation}
\Omega(r_h)=\sum\limits^m_{s=0}\tilde{\alpha}_s\kappa^s r^{-2s+n+1}_h+V(r_h)-\omega_0=0.\label{eq:rh}
\end{equation}
The quantity $\omega_0=\omega_c(\alpha_s,a)$ can be tuned to have $r_*(\alpha_s,a,\omega_0)=r_h(\alpha_s,a,\omega_0)\equiv r_0$. We consider only the $\kappa=1$ case here since a similar analysis can be easily applied for the other cases. At the critical point, Eq.~(\ref{eq:r*}) and Eq.~(\ref{eq:rh}) become
\begin{equation}
\sum\limits^m_{s=1}\tilde{\alpha}_ssr^{-2s+n+1}_0=0,\label{eq:r*0}
\end{equation}
and
\begin{equation}
\sum\limits^m_{s=0}\tilde{\alpha}_sr^{-2s+n+1}_0+V(r_0)-\omega_0 =0,\label{eq:rh0}
\end{equation} 
respectively.
These two equations lead to the condition $\omega_0=\omega_c(\alpha_s,a)$. Repeating the procedure of multiplying the lower order equation by $r^q$ and eliminating the the highest order term eventually reduces the set of equations to two linear equations: $ar+b=0,~cr+d=0$. The existence of a solution leads to $\omega_0=\omega_c(\alpha_s,a)$ and the condition will be $ad-bc=0$. We are interested in the behaviour in the neighbourhood of the greatest horizon radius $r_{hM}$. Deviation of $\omega_0$ from $\omega_c$ will separate $r_*$ and $r_h$ from $r_0$, giving $\Delta r_*$ and $\Delta r_h$. Varying Eq.~(\ref{Omega}) around $r_0$ we see how $r_*$ and $r_h$ are affected due to a change of $\omega_0$. The relation between $\Delta r_*$ and $\Delta f$ due to $\Delta\omega_0$ can be found to be
\begin{equation}
\left.\frac{d\Omega}{d r}\right|_0\Delta r_*+\frac{1}{2!}\left.\frac{d^2 \Omega}{d r^2}\right|_0\Delta r_*^2+\cdots-\Delta\omega_0 =0,
\end{equation}
where
\begin{equation}
\left.\frac{d\Omega}{d r}\right|_0=\lim\limits_{r\to r_0}\left(\frac{\partial \Omega}{\partial f}\frac{df}{dr}+\frac{\partial \Omega}{\partial r}\right). \label{DmOmega}
\end{equation}
The derivatives $\left.d^m\Omega/d r^m\right|_0$ are similarly obtained. 
In addition $d^mf/dr^m$ can be found by repeatedly taking derivatives of Eq.~(\ref{eq:r*}). For example, $df/dr$ can be found from
\begin{equation}
0=\frac{\partial^2 \Omega}{\partial f^2}\frac{df}{dr}+\frac{\partial^2 \Omega}{\partial r \partial f}, \cdots \label{Dmf}
\end{equation}

On the other hand, $\Delta r_h$ can be obtained by varying Eq.~(\ref{Omega}) with $f$ fixed, so that 
\begin{equation}
\left.\frac{\partial \Omega}{\partial r}\right|_0\Delta r_h+\frac{1}{2}\left.\frac{\partial^2 \Omega}{\partial r^2}\right|_0\Delta r_h^2+\cdots =\Delta\omega.
\end{equation}
We consider the example of $n=5$ with a string cloud background for the special choice of parameters, $\tilde{\alpha}_0=0$, $\tilde{\alpha}_1=1$ $\tilde{\alpha}^2_2=3\tilde{\alpha}_3\equiv\alpha^2$. This illustrates how $\omega$ changes $r_*$ and $r_h$ \cite{Lee:2014dha}. We have
\begin{equation}
\left\{
\begin{array}{lll}
&&-\frac{\partial\Omega}{\partial f}=r^4+2\alpha(1-f)r^2+(1-f)^2\alpha^2\\
&&\Omega=(1-f)r^4+\alpha(1-f)^2r^2+\alpha^2(1-f)^3/3
+\frac{2}{5}ar-\omega_0.
\end{array}\right.
\end{equation}
The first equation is factored out to be 
\begin{equation}
\frac{\partial\Omega}{\partial f}=-[r^2+\alpha(1-f)]^2, ~~ \left.\frac{\partial\Omega}{\partial f}\right|_0=0 \nonumber
\end{equation}
and gives $r_0=(-\alpha)^{1/2}$.  The critical point can be found from the second equation
\begin{equation}
\alpha^2+\frac{6}{5}a(-\alpha)^{1/2}-3\omega_0=0.\label{con:n5}
\end{equation}
We can calculate $d^pf/dr^p$ and $\partial^n\Omega/\partial r^p\partial f^q$ using from Eq.~(\ref{Dmf}) and Eq.~(\ref{DmOmega}), respectively
$\Delta \omega_0$ can be found as

\begin{equation}
(2r^3_0+2a/5)\Delta r_*+\frac{1}{2!}(10r^2_0)\Delta r^2_*+\frac{1}{3!}(40r_0)\Delta r^3_*+\cdots=\Delta \omega_0,
\end{equation}
and
\begin{equation}
(2r^3_0+2a/5)\Delta r_h+\frac{1}{2!}(10r^2_0)\Delta r^2_h+\frac{1}{3!}(24r_0)\Delta r^3_h+\cdots=\Delta \omega_0.
\end{equation}
If $r_0$ is the greatest horizon radius, then $\left.\frac{\partial \Omega}{\partial r}\right|_0=(2r^3_0+2a/5)>0$. Thus for $\Delta \omega_0>0$, $\Delta r_*>0$ and $\Delta r_h>0$. In these two equations coefficients start to differ in the third order with $40r_0>24r_0$. Expressing $\Delta \omega_0$ with $\Delta r_h=\Delta r_*+\delta$ we get 
\begin{equation}
\frac{1}{3!}(24r_0-40r_0)\Delta r^3_*+(a_1+2a_2\Delta r_*+3a_3\Delta r^2_*)\delta=0,
\end{equation}
where $a_1=(2r^3_0+2a/5)$, $a_2=\frac{1}{2!}(10r^2_0)$ and $a_3=\frac{1}{3!}(24r_0)$. It gives $\delta>0$ and hence yields $\Delta r_*<\Delta r_h$ for $\Delta \omega_0>0$. On the other hand $\Delta r_*>\Delta r_h$ for $\Delta \omega_0<0$.
With the metric function $f(r)=1+r/\alpha-g^{1/3}/\alpha$, where $g(r)=r^6-(6/5)a\alpha r+3\omega_0\alpha$, the above results can be verified.
It should be noted from the structures of $f(r)$ and $g(r)$, that increasing $\omega_0$ increases $r_h$ and $r_*$ around $r_0$, leading to $r_*<r_h$ and decreasing $\omega_0$ makes $r_*>r_h$.

\begin{center}
\textbf{Acknowledgements }
\end{center}
THL thanks the University of
KwaZulu-Natal for financial support.
SDM acknowledges that this work is based upon research supported by the South African Research
Chair Initiative of the Department of Science and
Technology and the National Research Foundation. 
SGG and DB acknowledges the NRF for research support.

\end{document}